\documentclass[10pt, conference]{IEEEtran}
\IEEEoverridecommandlockouts
\usepackage{amsmath,amssymb,amsfonts}
\usepackage{graphicx}
\usepackage[short]{optidef}
\usepackage{textcomp}
\usepackage{xcolor}
\usepackage{layout}
\def\BibTeX{{\rm B\kern-.05em{\sc i\kern-.025em b}\kern-.08em
    T\kern-.1667em\lower.7ex\hbox{E}\kern-.125emX}}

\begin{document}

\title{
Opportunistic Data Offloading for Robotic Operations in Dynamically Varying\\Channel Environments  
}
\author{Heeirthan Shanthan, Winston Hurst, Yasamin Mostofi
\thanks{Heeirthan Shanthan, Winston Hurst, Yasamin Mostofi are with the Department of Electrical and Computer Engineering, University of California, Santa Barbara, USA (email: \{heeirthan, winstonhurst, ymostofi\}@ece.ucsb.edu). This work was supported in part by NASA award 80NSSC25M7102.}
}

\maketitle

%%%%%%%%%%%%%%%%%%%%%%%%%%%%%%%%%%%%%%%%%%%%%%%%%%%%%%%%%%%%%%%%%%%%%%%%%%%%%%%%
\begin{abstract}
This paper studies energy-efficient operation of autonomous vehicles (AVs) in dynamic environments with moving obstacles and while communicating over mmWave channels. The obstacles induce severe attenuation of the mmWave channel resulting in a highly dynamic communication environment. In this setting, we consider the problem of jointly optimizing motion and communication energy for an AV that safely navigates among dynamic obstacles toward a designated destination while ensuring timely transmission of onboard sensing or telemetry data over mmWave channels. %under a delay constraint, modeled by a finite onboard data buffer. 
We then seek a real-time methodology to compute energy-efficient trajectories in a setting where dynamic obstacles induce both safety constraints and time-varying mmWave blockage, leading to tightly coupled motion–communication trade-offs.
We propose a nonlinear model predictive control (NMPC) framework that enables anticipative communication and motion decision-making and energy co-optimization, augmented with a control barrier function (CBF) to ensure safety. Extensive simulation results demonstrate the effectiveness of our approach, reducing total energy consumption by up to 37.3\% compared to baseline strategies. Overall, our results demonstrate that the proposed NMPC-based framework significantly enhances energy efficiency and performance of AVs under dynamic, blockage-sensitive mmWave communication constraints.

\end{abstract}
\vspace{-3pt}
%%%%%%%%%%%%%%%%%%%%%%%%%%%%%%%%%%%%%%%%%%%%%%%%%%%%%%%%%%%%%%%%%%%%%%%%%%%%%%%%
\section{Introduction}
\vspace{-3pt}
The ongoing development of standards for 6G wireless communication systems, coupled with advances in autonomous vehicles (AVs), has accelerated research into communication-aware robotics~\cite{c1, c2}. In many applications, AVs must maintain reliable communication with a remote operator or base station for command-and-control messaging, telemetry logging, or real-time data gathering. At the same time, AVs navigate increasingly dynamic environments, which require adaptive approaches to trajectory design.

To support the communication requirements of AVs, millimeter-wave (mmWave) communication, introduced in 5G and further developed for 6G systems, offers large bandwidths that support high data rates \cite{c3}. However, mmWave signals experience severe attenuation due to blockage, making reliable communication strongly dependent on maintaining line-of-sight (LOS) links between the transmitter and receiver \cite{c3}. In dynamic environments, such as outdoor facilities, moving obstacles, such as people or AVs moving through the area, can cause frequent changes between LOS/NLOS conditions, leading to highly varying channel quality and unreliable communication.

In this paper, we consider an AV navigating a dynamic environment while transmitting incoming data to a base station over a mmWave channel. We seek to jointly optimize the AV's trajectory and communication to minimize energy use while ensuring the timely offloading of the data. While prior work has explored co-optimization of motion and communication \cite{c15}-\cite{c17}, these approaches typically assume static environments or adopt communication models that do not capture the blockage-dominated and highly directional nature of mmWave links. In dynamic settings, moving obstacles induce abrupt channel variations, tightly coupling motion and communication decisions and necessitating new solutions.

To tackle this new challenge, we propose a real-time non-linear model predictive control (NMPC) based framework coupled with a high-frequency control barrier function (CBF) \cite{c10} for safe and energy-efficient operation of autonomous vehicles in mmWave environments with dynamic obstacles. The proposed approach jointly optimizes motion and communication energy, predicting NLOS regions caused by the dynamic obstacles and using adaptive transmission rates and trajectory planning to mitigate their effects. Extensive simulations in realistic communication environments confirm that our method can reduce total energy usage considerably (e.g., 37.3\%) and prevent buffer overflow, as compared to baseline methods, while successfully accomplishing the mission.

The rest of the paper is organized as follows. In Section~\ref{sec:modeling}, we present the AV and buffer dynamics, along with the communication and channel models. Section~\ref{sec:probform} introduces the general problem formulation and its associated challenges. In Section~\ref{sec:sysov}, we reformulate the problem and present the proposed solution approach. Finally, Section~\ref{sec:results} presents simulation results, followed by concluding remarks in Section~\ref{sec:conc}. 

\begin{figure}
\centering
\includegraphics[width=0.75\linewidth]{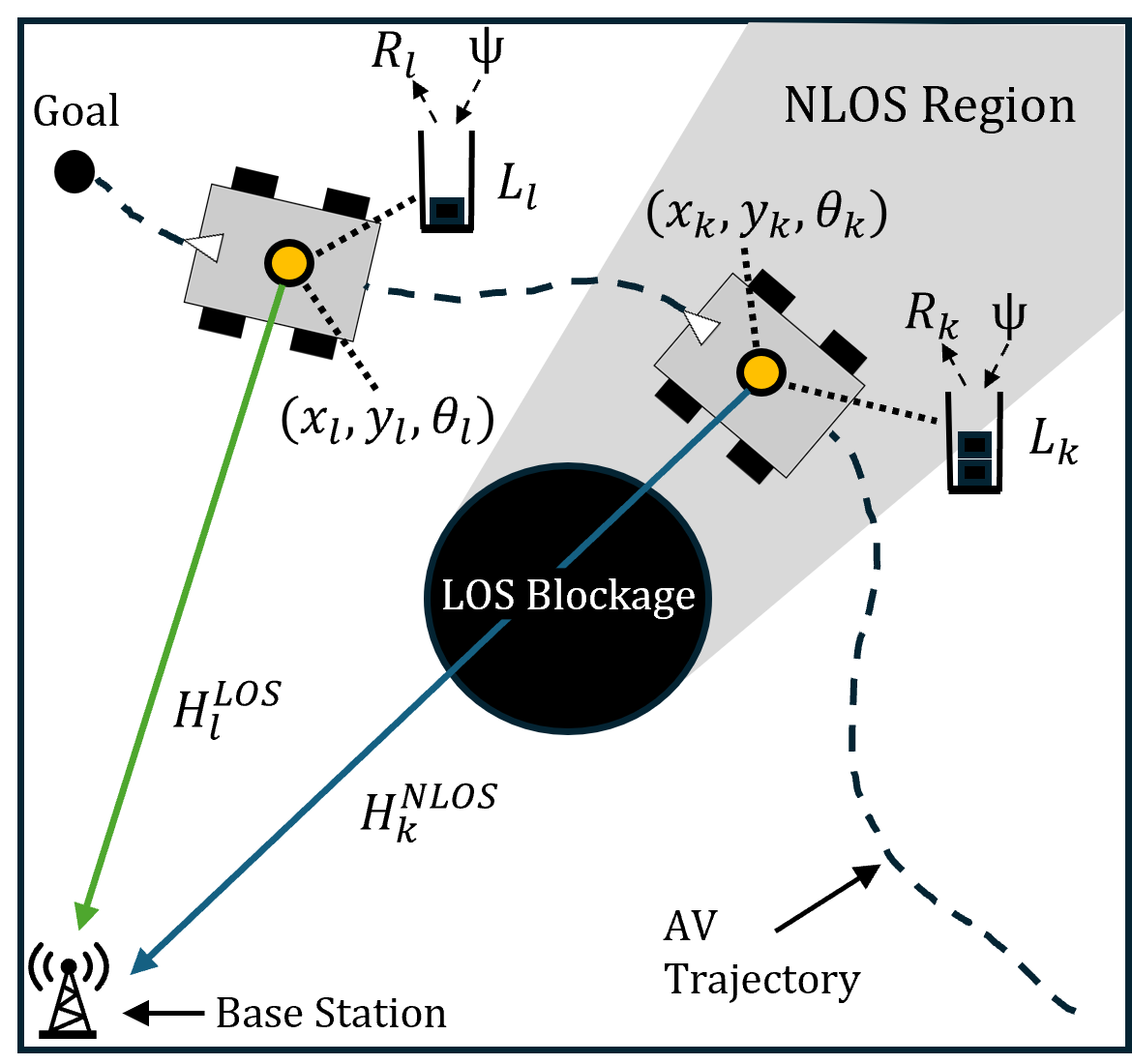}
\vspace{-10pt}
\caption{Overview of the problem considered in this paper. An AV has to navigate to its destination while avoiding dynamic obstacles, such as pedestrians and other AVs (black circle). At time step \(k\), the AV, at position \(x_k,y_k\) with heading \(\theta_k\), has an onboard finite-length data buffer containing \(L_k\) bits of data. The AV collects data in its buffer at a constant rate \(\psi\), and offloads data to the base station at a selected rate \(R_k\). Dynamic obstacles can block the direct LOS between the AV and the base station, creating NLOS regions that result in increased attenuation of the channel \(H_k\).}
\vspace{-20pt}
\label{fig:scenario}
\end{figure}
\vspace{-10pt}
\section{System Modeling}\label{sec:modeling}
\vspace{-8pt}
Consider the scenario shown in Figure~\ref{fig:scenario}, where an AV must navigate from an initial position to a target destination in a dynamic environment with moving obstacles. During its operation, data from the sensors on the AV accumulates in an onboard memory buffer and shall be offloaded in a timely manner to a nearby base station over mmWave communication channels. We next present a mathematical model of the system.

\subsection{Onboard Buffer Dynamics}
The AV collects data at a constant rate of $\psi$ bps and offloads at a rate of $R_k = Br_k$ bps, where $B$ is the communication bandwidth and $r_k$ is the spectral efficiency in bps/Hz. Let $L_{k}$ denote the buffer length at time index $k$, with a maximum capacity of $L_{max}$. 
The buffer dynamics are described by 
\vspace{-5pt}
\begin{equation}
     L_{k+1} = \big( L_k + (\psi - R_k)\Delta t \big),
     \vspace{-5pt}
\end{equation}
where $\Delta t$ is the duration of a single time step. %and \(\mathrm{sat}_{[0,L_{\text{max}}]}(L) = \max\big(0, \min(L_{\text{max}}, L)\big)\). %The saturation function ensures that the buffer length remains within the admissible range \([0,L_{\text{max}}]\).
The buffer length is restricted to the admissible range \([0,L_{\text{max}}]\).
The upper bound captures the finite storage capacity of the buffer, while the lower bound ensures that the buffer length remains non-negative. Alternatively, given the constant incoming data rate, the upper bound can be seen as a constraint on the maximum age of information \cite{c19}, i.e., the time between when the data is generated and when it is sent to the base station.
\vspace{-6pt}
\subsection{Communication and Channel Model}

Buffer dynamics depend on the transmission rate, which is determined by the wireless channel conditions. We next introduce our communication model.
The channel power between the AV and the base station at time step $k$, corresponding to the squared magnitude of the complex-valued baseband channel, is denoted by $H_k$.  Let $\sigma^2$ represent the receiver noise power. The resulting channel-to-noise ratio (CNR) is given by \(\gamma_{k} = H_k/\sigma^2\). We assume the commonly used MQAM modulation for transmission from the AV to the base station, where the quality of service (QoS) requirement is specified through a maximum acceptable bit error rate (BER), $p_{\text{BER}}$. Under this assumption, the minimum required transmission power to achieve a spectral efficiency of $r$~bps/Hz is well approximated by \(P_{T} = (2^r - 1)\ln({5p_{\text{BER}})}/(-1.5\gamma)\) \cite{c4}.
We consider a mmWave communication channel, which suffers from acute penetration loss. The channel model consists of path loss and small-scale multipath fading. Let $d$ denote the distance between the base station and the AV. The path loss in dB is modeled using a slope-intercept model: \(H_{\text{PL},\text{dB}} = C_{PL}\ -10\eta\log_{10}(d)\). Small-scale fading is modeled as a Rician random variable with Rician factor $K$. We consider two distinct sets of model parameters \(C_{\text{PL}},\eta,K\) depending on NLOS/LOS conditions. The LOS parameters result in a stronger CNR compared to the NLOS parameters, reflecting the attenuation caused by blockage in the environment. 

\subsection{Dynamic Model}
The AV is modeled using a discrete-time model for a differential drive robot. The AV state at time step $k$ consists of the planar position \((x_k,y_k)\), heading $\theta_k$, and buffer length $L_k$, while control inputs to the system are linear velocity $v_k$, angular velocity $\omega_k$, and offloading rate $R_k$. The resulting state and control vectors are
\vspace{-5pt}
\begin{equation}
\begin{aligned}
\mathbf{z}_k &= 
\begin{bmatrix}
x_k\;\; y_k\;\;\theta_k\;\;L_k
\end{bmatrix}^T,
&
\mathbf{u}_k &= 
\begin{bmatrix}
v_k \;\;\omega_k\;\;R_k
\end{bmatrix}^T.
\end{aligned}
\vspace{-5pt}
\end{equation}
The system dynamics are given by 
\vspace{-5pt}
\begin{equation}
f(\mathbf{z_{k}},\mathbf{u_{k}}) =
\begin{bmatrix}
x_k + v_k\cos(\theta_k)\,\Delta t \\
y_k + v_k\sin(\theta_k)\,\Delta t \\
\theta_k + \omega_k\,\Delta t \\
L_k + (\psi - R_k)\Delta t\,
\end{bmatrix}.
\vspace{-5pt}
\end{equation}

Although acceleration is not directly used as a control input, acceleration bounds are enforced by imposing rate constraints on the control inputs between successive time steps.
The input and rate constraints in discrete time are
\vspace{-5pt}
\begin{equation}
\begin{aligned}
\mathbf{u}_{\min} \le \mathbf{u}_k \le \mathbf{u}_{\max}, \quad
{\mathbf{a}}_{\min}\Delta t \le \mathbf{u}_{k+1} - \mathbf{u}_k \le {\mathbf{a}}_{\max}\Delta t\,,
\end{aligned}
\vspace{-5pt}
\end{equation}
applied element-wise to inputs $v$, $\omega$, and $R$.

Furthermore, we assume accurate measurements of the AV's states at the current time step and use the same dynamic model for prediction in the optimization framework. 
\vspace{-3pt}
\subsection{Dynamic Obstacles}
\vspace{-3pt}
The dynamic obstacles represent moving agents in the environment, such as pedestrians or other AVs operating within a structured setting (e.g., an outdoor facility). The AV’s onboard sensors provide obstacle position estimates at the current time step, which are assumed to be accurate.
Future obstacle motion is then predicted using a deterministic internal model for the purpose of optimization in the next section. We note that the true obstacle positions can deviate from this prediction, as we shall consider in Section~\ref{sec:results}. 
\vspace{-7pt}
\section{Problem Formulation}\label{sec:probform}
\vspace{-3pt}
Consider an AV navigating from an initial state \(z_{\text{init}}\) to a final state \(z_f\) in a 2D environment, while avoiding dynamic obstacles that occupy the set \(D\). The AV is equipped with an onboard data buffer whose length must remain within the interval \([0,L_{\text{max}}]\). To achieve this, the AV must compute motion and communication decisions over time. At each time step $k$, in addition to selecting appropriate velocity commands, the AV must choose an offloading rate \(R_k\) to ensure timely data offloading and prevent buffer overflow due to data accumulation from the AV's sensors at a constant rate $\psi$. The AV's transmit power, given by \(P_{T,k} = (2^{R_{k}/B} - 1)\ln({5p_{\text{BER}})}/(-1.5\gamma_{k})\), must not exceed a maximum allowable value \(P_{T,\text{max}}\). By inverting this expression, the constraint gives an upper bound on the achievable transmission rate:
\vspace{-5pt}
\[
R_k \le R^P_{\text{max}}(\gamma_{k}) =
B\log_2 \left(
1 + \frac{-1.5 \gamma_{k} P_{T,\max}}{\ln(5p_{\text{BER}})}
\right).
% \vspace{-5pt}
\]
% \enlargethispage{-0.1in}
In regions with high CNR, this bound may exceed hardware limitations. To account for this, we impose an additional constraint based on the maximum physically achievable transmission rate of the AV, denoted by $R^{\text{HW}}_{\text{max}}$, which is incorporated into the control limits \([u_{\text{min}},u_{\text{max}}]\). 

In this setting, the objective is to minimize the motion and communication energy of the AV, while achieving timely arrival at the destination and ensuring prompt data offloading. We propose the following general optimization problem:
\begin{mini}[2]
    { \{u_k, T\} } 
    {
    \begin{aligned}
        \sum_{k=0}^{T-2} \ell_{a,k} + \sum_{k=0}^{T-1} \big( \ell_{u,k} + \ell_{\text{comm},k} \big) + T 
    \end{aligned}
    }{\label{equationGeneral}}{}
\addConstraint{z_0}{= z_{\text{init}}, \quad z_T = z_f}{}  
\addConstraint{z_{k+1}}{= f(z_k, u_k),}{\quad k = 0,\dots,T-1 }
\addConstraint{(x_k, y_k)}{\notin D_k,}{\quad k = 0,\dots,T}
\addConstraint{u_{\min}}{\le u_k \le u_{\max},}{\quad k = 0,\dots,T-1}
\addConstraint{a_{\min}}{\le \frac{u_{k+1} - u_k}{\Delta t} \le a_{\max},}{\quad k = 0,\dots,T-2}
\addConstraint{R_k}{\leq R^{P}_{\max},}{\quad k = 0,\dots,T-1}
\addConstraint{0}{\le L_k \le L_{\max},}{\quad k = 0,\dots,T.}
\end{mini}
where $T$ denotes the final time step. The motion-related cost terms $\ell_{u,k} = u_k^\top W u_k$ and $\ell_{a,k} = (u_{k+1} - u_k)^\top W_a (u_{k+1} - u_k)$ penalize control effort and acceleration, respectively. The weighting matrices \(W\) and \(W_a\) assign zero weight to the transmission rate \(R_k\), so that only the kinematic components are penalized. The communication cost term $\ell_{\text{comm},k} = \delta P_{T,k}$ penalizes the transmission power used for data offloading, where \(\delta\) is a weighting coefficient. The weights are tuned to adjust the relative importance of motion and communication energy, as well as prompt arrival at the destination.

The main challenge arises from the trade-offs associated with motion-communication co-optimization. A communication-efficient strategy may favor trajectories that deviate toward regions of high CNR, while a motion-efficient strategy may favor the shortest feasible path, leading to conflicting objectives. Additional complexity is introduced by the nonconvex AV dynamics and the possibly nonconvex constraint \((x_k,y_k) \notin D_k\), as well as the objective of ensuring timely arrival at the destination and the need for real-time implementation. As a result, the problem is quite challenging to solve directly, especially in mmWave communication environments, motivating our proposed approach in Section~\ref{sec:sysov}.
\vspace{-5pt}
\section{NMPC-CBF Optimization Framework}\label{sec:sysov} 
\vspace{-3pt}
We next develop a novel NMPC-based approach to solve an efficient approximation of the optimization problem in~\eqref{equationGeneral}. Nonlinear model predictive control (NMPC) is well suited for constrained nonlinear optimal control, enabling non-myopic decision making. The general formulation involves a free terminal time $T$, which may lead to overly slow trajectories, increase the complexity of real-time implementation, and limit compatibility with finite-horizon NMPC. To address this, we instead introduce a reference path, computed prior to the AV mission as the shortest path between the initial and final positions, used as a tracking objective in the NMPC to promote timely progress toward the destination and reduce the search space of the optimization. For simple scenarios, this path reduces to a straight line, while in more complex environments it can be generated using A* as in~\cite{c7}. 
% \enlargethispage{-0.1in}
During execution, the NMPC computes motion and transmission rate commands by solving an optimal control problem over a finite prediction horizon at each time step, using predictive system models. The first control input of the resulting sequence is used as the NMPC output. To provide safety guarantees with respect to dynamic obstacles, we incorporate a control barrier function (CBF) as a safety filter that minimally modifies the NMPC output when necessary to ensure that the AV remains outside obstacle safety margins. The resulting control is applied to the system, and the updated state and obstacle measurements are used to repeat the process at the next time step.
\vspace{-3pt}
\subsection{NMPC Formulation}
We next present an NMPC formulation of the problem defined in Section~\ref{sec:probform}. Over a prediction horizon of length $N$, we consider a sequence of control inputs $\{u_j\}_{j=0}^{N-1}$ that minimizes
$J = \sum_{j=0}^{N-1} \ell(z_j, u_j) + \ell_f(z_N)$,
where $\ell(\cdot)$ denotes the stage cost, $\ell_f(\cdot)$ the terminal cost, and $j$ indexes the prediction steps within the horizon. At time step $k$, $z_k$ denotes the current system state and initializes the prediction with \(z_0=z_k\).
The resulting optimal control problem is:
\begin{mini}[2]
    { \{u_j\} } 
    {
    \begin{aligned}
    &\sum_{j=0}^{N-2} \ell_{a,j}
    + \sum_{j=0}^{N-1} (\ell_{u,j} + \ell_{z,j})\ + \\
    & \sum_{j=0}^{N-1} (\ell_{g,j} + \ell_{\text{comm},j})
    + \ell_{z,N} + \ell_{g,N}
    \end{aligned}
    }{\label{equationNMPC}}{}
\addConstraint{z_0}{= z_k}{}
\addConstraint{z_{j+1}}{= f(z_j, u_j),}{\quad j = 0,\dots,N-1}
\addConstraint{(x_j, y_j)}{\notin D_j,}{\quad j = 0,\dots,N}
\addConstraint{u_{\min}}{\le u_j \le u_{\max},}{\quad j = 0,\dots,N-1}
\addConstraint{a_{\min}}{\le \frac{u_{j+1} - u_j}{\Delta t} \le a_{\max},}{\quad j = 0,\dots,N-2}
\addConstraint{R_j}{\leq R^{P}_{\max},}{\quad j = 0,\dots,N-1}
\addConstraint{0}{\le L_j \le L_{\max},}{\quad j = 0,\dots,N.}
\end{mini}
The cost terms $\ell_{z,j} = (z_j - z_j^{\mathrm{ref}})^\top Q (z_j - z_j^{\mathrm{ref}})$ and
$\ell_{z,N} = (z_N - z_N^{\mathrm{ref}})^\top Q_N (z_N - z_N^{\mathrm{ref}})$ penalize deviation from the corresponding waypoints $z_j^{\mathrm{ref}}$ and $z_N^{\mathrm{ref}}$, extracted from the reference path. Strictly following the reference path may not be optimal from a communication perspective, as the AV may benefit from deviating towards regions with higher CNR \(\gamma_j\). To allow such deviations while maintaining timely progress toward the destination, we introduce $\ell_{g,j} = (z_j - z_f)^\top Q_g (z_j - z_f)$ and $\ell_{g,N} = (z_N - z_f)^\top Q_{g,N} (z_N - z_f)$, which penalize distance to the final state. The weighting matrices for \(\ell_z\) and \(\ell_g\) assign zero weight to the buffer state $L$, such that only the kinematic components are penalized. By tuning the relative weights of the cost terms, the controller can balance motion, communication, and arrival time objectives. At each time step $k$, a control sequence is computed, and only the first control input is used. The optimization is then repeated at the next time step using updated system measurements. 
To solve the resulting NMPC problem in real-time, we employ the PANOC algorithm \cite{c9}, which is well suited for real-time application due to its computational efficiency and ability to handle nonlinearities and nonconvexities.
\vspace{-6pt}
\subsection{Prediction Models}
\vspace{-5pt}
NMPC requires models describing the system evolution over the prediction horizon. The AV state is assumed to be accurately measured at the current time step $k$, with future states predicted using the AV dynamics. The obstacle positions are assumed to be accurately measured at $k$, and their future positions are predicted using a deterministic linear motion model. In practice, obstacle positions may naturally deviate from this model, which is accounted for in the simulations. For the channel model, the path loss component is computed directly from the predicted AV-base station distance using the corresponding LOS/NLOS parameters. In contrast, small-scale fading power varies rapidly over small spatial regions and is difficult to predict, as shown in the literature. Therefore, we use the expected channel gain in the prediction model. Since the Rician fading distribution is normalized such that the average small scale fading power equals 1, the expected channel gain simplifies to \(\mathbb{E}[H] = H_{\text{PL}}\), i.e., it depends only on path loss and the LOS/NLOS condition. While the NMPC relies on this expected model for tractability, the simulation results use stochastic small-scale fading realizations to evaluate performance under realistic channel variability. In practice, LOS/NLOS parameters for mmWave channels can be learned from a small set of prior channel measurements, e.g., gathered at the beginning of the operation, offline, or via crowdsourcing~\cite{c20}. In this work, we adopt experimentally validated parameters.
\vspace{-3pt}
\subsection{CBF Safety Filter}
\vspace{-2pt}
The NMPC does not provide formal safety guarantees and may allow violations of obstacle safety margins due to real-time implementation and solver tolerances. Therefore, we incorporate a CBF as a safety filter~\cite{c10} to enforce safety.

Consider a control-affine system of the form \(\dot{z} = f(z) + g(z)u\),
where $z$ denotes the system state and $u$ is the control input. The objective is to compute a control input that renders the safe set $\mathcal{A} = \{ z \mid h(z) \ge 0 \}$, defined by the control barrier function $h(z)$, forward invariant while staying close to the nominal control input generated by the NMPC. This is achieved by solving the following quadratic program (QP):
\begin{equation}
\begin{aligned}
\min_{u} \quad & \|u - u_{\text{nom}}\|_2^2 \\
\text{s.t.} \quad & L_f h_2(z) + L_g h_2(z)u \ge -\alpha_2\big(h_2(z)\big),
\end{aligned}
\end{equation}
where \(h_2(z) = L_f h(z) + \alpha_1\big(h(z)\big)\), \(u_{\text{nom}}\) is the nominal control input generated by the NMPC, and \(\alpha_1(\cdot)\), \(\alpha_2(\cdot)\) are extended class \(\mathcal{K}_\infty\) functions. Further details on the CBF formulation can be found in~\cite{c10}. To enforce collision avoidance with circular dynamic obstacles, we define \(h(z) = (x-x_0)^2 + (y-y_0)^2 - r_{\text{safe}}^2\)
% \begin{equation}
% \vspace{-5pt}
% h(z) = (x-x_0)^2 + (y-y_0)^2 - r_{\text{safe}}^2 
% \end{equation}
where $(x_0,y_0)$ denotes the obstacle center and $r_{\text{safe}}$ defines the safety margin. 

\begin{figure*}[t]
    \centering
    \includegraphics[width=0.93\textwidth]{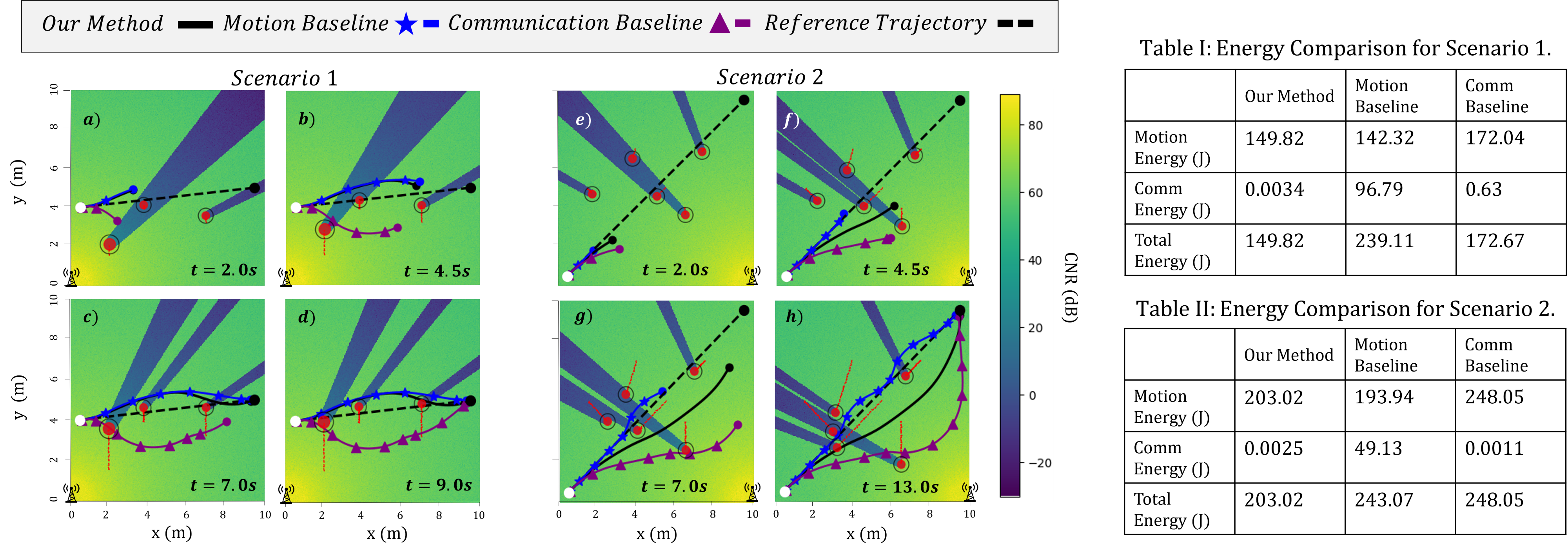}
    \vspace{-15pt}
    \caption{Trajectories for all methods in Scenario 1 (a-d) and Scenario 2 (e-h). The subplots show snapshots at successive time instances of the simulation. Brighter (darker) colors in the colormap indicate higher (lower) CNR. The white and black dots denote the initial and final AV positions, respectively. Red circles indicate the dynamic obstacles, with black circles as their safety regions \(r_{\text{safe}}\). Red dashed lines show obstacle trajectories. The base stations are located at (0,0) for Scenario 1 and (10,0) for Scenario 2. Note that even when our trajectories become close to motion baseline, the transmission strategy differs substantially, yielding significant energy savings, as can be seen from Table I and II. See the color PDF for optimal viewing.}   
    \vspace{-19pt}
    \label{fig:casestraj}
\end{figure*}
\enlargethispage{-0.1in}
\vspace{-9pt}
\section{Simulation Results}\label{sec:results}
In this section, we evaluate our approach through a series of simulations. We first examine the motion and communication energy consumption for two different scenarios, comparing against two baselines with limited communication awareness. These studies indicate that our method can significantly lower the total energy, particularly when NLOS regions are unavoidable. We then study the impact of the incoming data rate, \(\psi\), on the trajectories generated by our framework and show that our method adapts both communication scheduling and route planning based on communication needs. In all, our method greatly enhances operation through an adaptive transmission rate and predictive trajectory planning in dynamic mmWave channels.

We consider two NMPC-based methods as natural baselines for our method. In the \textbf{motion baseline}, the AV optimizes only for motion efficiency, without considering the communication cost terms or constraints in \eqref{equationNMPC}, while data is transmitted at a fixed rate \(R=\psi\). In the \textbf{communication baseline}, both motion and communication objectives and constraints are considered in the optimization, as the problem formulation matches \eqref{equationNMPC}. However, the transmission rate remains fixed at \(R=\psi\). 
As a result, communication energy can only be reduced by moving towards regions of strong CNR, forcing the AV to avoid NLOS regions. It is noteworthy that while the motion baseline aligns with standard state-of-the-art approaches, the communication baseline already embeds some of the components of our proposed framework, though it lacks the critical capability of adaptive rate optimization.
\vspace{-5pt}
\subsection{Simulation Setup}
For all scenarios, we consider a \(10\text{m} \times 10\)m 2D workspace. Obstacle motion is described by a deterministic linear dynamical model, but with additive Gaussian white noise to capture realistic deviations from the nominal behavior. The BER used for determining the quality of service (QoS) requirement is \(p_{\text{BER}}=10^{-6}\). The receiver noise power is \(\sigma^2=-85\)~dBm, and the channel bandwidth is 100 MHz \cite{c18}. The onboard buffer is initially loaded with \(L_0 = 400\)~Mb, has maximum capacity \(L_{\text{max}}=700\)~Mb, with an incoming data rate of \(\psi = 500\)~Mbps. Based on experimental measurements at 28~GHz \cite{c6}, we set the NLOS/LOS parameters for the channel model to \(\eta_{\text{NLOS}}= 5.76\), \(\eta_{\text{LOS}}= 2.5\), \(C^{\text{LOS}}_{\text{PL}} = 0\)~dB, \(C^{\text{NLOS}}_{\text{PL}} = -25\)~dB, \(K_{\text{NLOS}} = 7\)~dB, and \(K_{\text{LOS}} = 12\)~dB. The maximum allowable transmission power is \(P_{T,\text{max}} = 1\)~W. 

The sampling interval is \(\Delta t=0.1\)s. The weighting matrices are chosen as \(Q=\mathrm{diag}(0.1,0.1,0.0,0.0)\) and \(Q_{N}=\mathrm{diag}(5.0,5.0,0.0,0.0)\), with \(Q_g = Q\) and \(Q_{g,N} = Q_N\), where \(\mathrm{diag}(\cdot)\) denotes a diagonal matrix.  Moreover, the communication weight is set to \(\delta = 10.0\). Overall, these weights strike a balance between communication and motion objectives.  The input and acceleration costs are weighted by $W = \mathrm{diag}(0.5,0.1,0.0)$ and $W_a = \mathrm{diag}(0.1,0.1,0.0)$, respectively, which are also used for the baselines to ensure a fair comparison. The CBF safety filter is realized using the high-frequency framework CBFpy \cite{c12}. The obstacle safety radius $r_{\text{safe}}$ is set to 0.35~m. For Scenario 1, we consider an NMPC horizon of \(N=30\) and for Scenario 2, \(N=40\).
To evaluate motion energy, we adopt the discrete time motor energy consumption model for a Pioneer 3DX robot in \cite{c14}:
\vspace{-6pt}
\begin{equation*}
\begin{split}
E_{\text{motor}} =
\sum_k \Big(
m \max\{v_k a_k, 0\}
+ I \max\{\omega_k \beta_k, 0\} \\
+ 2\mu mg \max\{|v_k|, |b\omega_k|\}
\Big)\,\Delta t,
\end{split}
\label{equationME}
% \vspace{-5pt}
\end{equation*}
where \(a_k\) and \(\beta_k\) denote the linear and angular accelerations of the AV, respectively. The parameter $m$ is the mass of the robot, $I$ is its moment of inertia, $\mu$ is the rolling friction coefficient, $g$ denotes the gravitational acceleration, and $b$ represents the robot radius. The parameter values are chosen according to \cite{c14} as \(m = 9\,\text{kg}\), \(I = 0.162\,\text{kg\,m}^2\), \(\mu =0.082\), \(b = 0.185\)~m, and \(g = 9.81\,\text{m/s}^2\). Finally, the input constraints and rate limits are defined as
\vspace{-5pt}
\begin{equation*}
\begin{aligned}
u_{\min} &= [-0.5\;\;-0.5\;\;0.0]^T, \quad
u_{\max} = [1.5\;\;0.5\;\;1.2]^T, \\
a_{\min} &= [-1.0\;\;-3.0\;\;-1.0]^T, \quad
a_{\max} = [1.0\;\;3.0\;\;1.0]^T.
\end{aligned}
\end{equation*}
The constraints and rate limits for \(v\) and \(\omega\) are given in \(\mathrm{m/s}\), \(\mathrm{m/s^2}\), \(\mathrm{rad/s}\), and \(\mathrm{rad/s^2}\), respectively, while those for \(R\) are in \(\mathrm{Gbps}\) and \(\mathrm{Gb/s^2}\).
\vspace{-5pt}
\subsection{Comparison with Baseline Methods}
\vspace{-2pt}
We first consider Scenario 1 in which three dynamic obstacles move across the nominal reference path, making it difficult for the robot to avoid NLOS regions. Figure~\ref{fig:casestraj} (a-d) shows the trajectories generated by our method and the two baselines, while Table I shows that our method lowers the total energy by 37.3\% and 13.2\% relative to the motion and communication baselines, respectively, and reduces communication energy by orders of magnitude. 

\begin{figure}[t]
\centering
\includegraphics[width=0.75\linewidth]{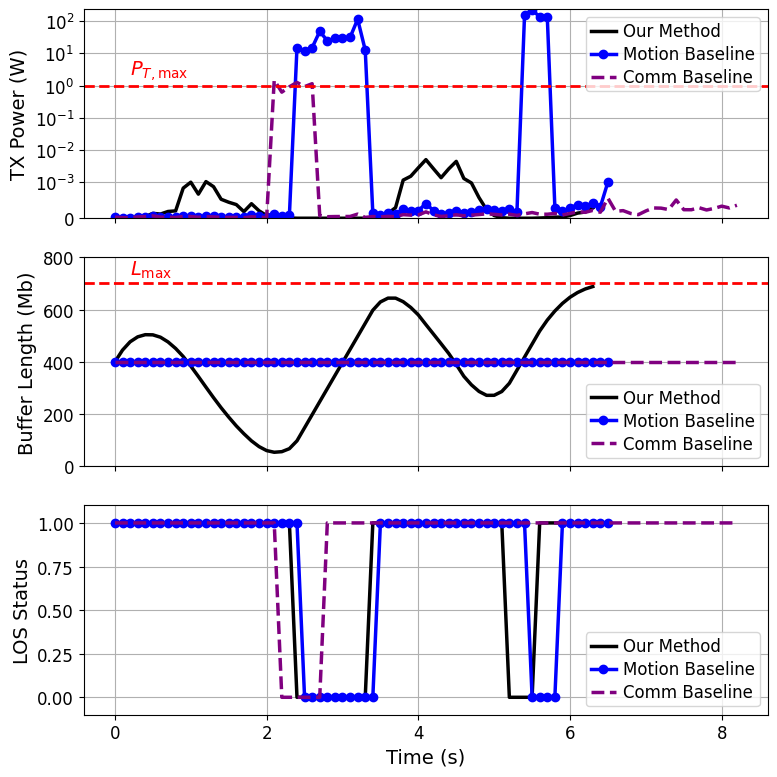}
\vspace{-14pt}
\caption{Simulation results for Scenario 1. For all three methods, the subplots show: (top) transmission power, (middle) buffer length, and (bottom) LOS status (1 for LOS, 0 for NLOS) over time. Note that the plots terminate at different times due to varying trajectory durations. See the color PDF for optimal viewing.}
\vspace{-13pt}
\label{fig:case1buff}
\end{figure}

Comparing our approach to the motion baseline, both trajectories remain relatively close to the reference path in this case and therefore enter three NLOS regions (due to the makeup of the environment as well as the value of $\psi$). However, while the motion baseline continues to transmit in these regions, it requires transmission powers exceeding the maximum allowable limit \(P_{T,\text{max}}=1\)W by a significant margin, as seen in Figure~\ref{fig:case1buff} (top). Our controller anticipates upcoming NLOS regions and empties the buffer in advance, thereby avoiding data transmission in NLOS conditions, as illustrated in Figure~\ref{fig:case1buff} (middle). Additionally, as seen in Figure~\ref{fig:casestraj} (b-c), our trajectory detours to exploit an LOS region and relieve the buffer before reaching the destination, a behavior not observed in the motion baseline.

On the other hand, the communication baseline depends highly on LOS conditions. As shown in Figure~\ref{fig:casestraj}, the communication baseline selects a detour that incurs higher motion energy in order to avoid the second and third NLOS regions. In contrast, our method avoids data transmission in the NLOS regions by properly adapting the rate in the LOS regions, which avoids the need for costly detours.  

In Scenario 2, five dynamic obstacles move in the space of interest. Figure~\ref{fig:casestraj} (e-h) shows the three very distinct trajectories (our method and the two baselines), while Table II shows that our method reduces the total energy consumption by 16.5\% and 18.2\% relative to the motion and communication baselines, respectively.
While the motion baseline passes through multiple NLOS regions (costly for communication), the communication baseline avoids all NLOS regions by moving closer to the base station (costly for motion). In contrast, our approach balances the trade-off between large trajectory deviations and traversing NLOS regions by finding a trajectory that limits exposure to a single NLOS region with only a minor detour from the reference trajectory. Furthermore, our approach avoids transmission in the NLOS region by preemptively emptying the buffer, similar to Scenario 1.
\vspace{-6pt}
\subsection{Impact of Data Accumulation Rate}
\begin{figure}
\centering
\includegraphics[width=0.75\linewidth]{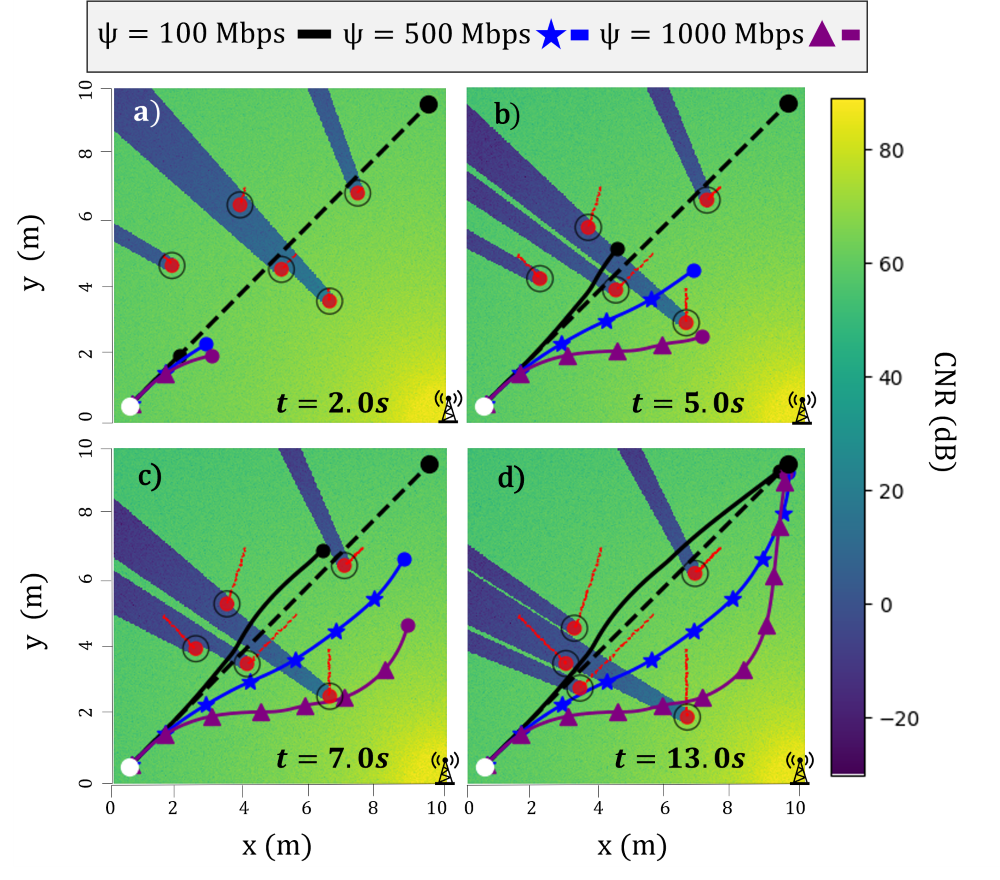}
\vspace{-13pt}
\caption{Trajectories for three different incoming data rates \(\psi\) for Scenario 2. Subplots (a-d) show snapshots at successive time instances of the simulation. Plot elements and color coding are consistent with Figure~\ref{fig:casestraj}, including the reference path (black, dashed line). See the color PDF for optimal viewing.}
\vspace{-16pt}
\label{fig:psi_traj}
\end{figure}

\vspace{-3pt}
We next study the impact of the incoming data rate \(\psi\) on the behavior of our method. Figure~\ref{fig:psi_traj} shows the adaptability of our method, where the trajectories vary with $\psi$. For a low rate, \(\psi=100\)~Mbps, the AV does not avoid any of the NLOS regions, since the low incoming rate allows it to avoid transmission in these regions without risking buffer overflow. For a moderate rate, \(\psi=500\)~Mbps, the behavior matches the main result for Scenario 2, where the robot allows limited exposure to NLOS regions while avoiding large trajectory deviations. For a high rate \(\psi=1000\)~Mbps, the AV detours significantly to avoid NLOS regions entirely, as pausing transmission even briefly will result in buffer overflow.  
\enlargethispage{-0.05in}
\vspace{-4pt}
\subsection{Computational Efficiency}
% \vspace{-3pt}
Our method is well suited for real-time applications. Average solver times is 34 ms and 55 ms per step for Scenario 1 and 2, respectively, well below the 100 ms sampling interval.
\vspace{-4pt}
\subsection{Impact of Safety Filter (CBF)}
% \vspace{-3pt}
The CBF provides safety guarantees that are not ensured by the NMPC alone, particularly due to numerical solver tolerances and real-time computation limits. For instance, when Scenario 2 is under a high incoming data rate of \(\psi=1000\)~Mbps, removing the CBF results in a collision with the rightmost obstacle. Thus, the CBF is critical for ensuring safe operation while preserving real-time feasibility.
\vspace{-6pt}
\section{Conclusions}\label{sec:conc}
\vspace{-3pt}

In this paper, we presented an NMPC-based framework with a CBF safety filter to handle collision-free adaptive trajectory design and data offload scheduling in dynamic mmWave communication environments induced by moving obstacles. Our method achieves real-time, predictive decision-making, reducing total energy consumption by up to $37.3$\% compared to baseline strategies in simulated environments. Additionally, our method shows orders of magnitude reduction in communication energy when NLOS regions are unavoidable. Overall, our method greatly enhances the operation of AVs utilizing mmWave communication channels.
\addtolength{\textheight}{-12cm}

\end{document}